# Superconducting magnesium diboride films with $T_c \approx 24$K grown by pulsed laser deposition with *in-situ* anneal

H.M. Christen, H.Y. Zhai, C. Cantoni, M. Paranthaman, B.C. Sales, C. Rouleau,
D.P. Norton*, D.K. Christen, and D.H. Lowndes

Oak Ridge National Laboratory, Oak Ridge, TN 37931-6056
*University of Florida, PO Box 116400, Gainesville FL 32611-6400

Thin superconducting films of magnesium diboride ($MgB_2$) with $T_c \approx 24$K were prepared on various oxide substrates by pulsed laser deposition (PLD) followed by an *in-situ* anneal. A systematic study of the influence of various *in-situ* annealing parameters shows an optimum temperature of about 600°C in a background of 0.7 atm. of Ar/4%H$_2$ for layers consisting of a mixture of magnesium and boron. Contrary to *ex-situ* approaches (e.g. reacting boron films with magnesium vapor at $\approx$ 900°C), these films are processed below the decomposition temperature of $MgB_2$. This may prove enabling in the formation of multilayers, junctions, and epitaxial films in future work. Issues related to the improvement of these films and to the possible *in-situ* growth of $MgB_2$ at elevated temperature are discussed.

The recent discovery of superconductivity at 39K in the simple binary compound $MgB_2$ [1,2] has resulted in numerous research efforts concentrating on optimizing and understanding physical properties [3] and synthesis [4]. Speculations about potential applications are motivated by the simplicity of the crystal structure of $MgB_2$ (as compared to the copper-oxide based high-temperature superconductors), the low cost of the starting materials, and the fact that grain boundaries appear to have a less detrimental effect on the superconducting current transport than in high-temperature superconductors [5,6]. However, the growth of $MgB_2$ films is complicated by the large differences in vapor pressure (and possibly adsorption coefficients and mobilities) between boron and magnesium.

Currently the films with the highest transition temperatures have been achieved by an *ex-situ* process in which boron films are reacted with magnesium vapor at temperatures near 900°C, as reported by us and other groups [7,8], and similar to the approach used to fabricate $MgB_2$ wires [4]. This approach, however, requires a heat treatment above the decomposition temperature of $MgB_2$ and is thus not applicable to the growth of multilayers and junctions. Films resulting from this method also exhibit rough surfaces. Furthermore, the conversion of a boron film via annealing in a Mg vapor requires the magnesium has to diffuse through an already formed $MgB_2$ surface layer. The portion of the film in contact with the substrate reacts last, thus making it very difficult to obtain an epitaxial relationship between the film and the substrate.

The films discussed here either contain the correct amount or an excess of magnesium before the annealing process, which may facilitate the nucleation of oriented grains or epitaxial films in future work, and significantly reduces the temperature required to form $MgB_2$. Clearly, the present approach could also be applied to other methods including co-sputtering or co-evaporation.

The experimental set-up used for this work consisted of a conventional PLD system with a base pressure of $5 \times 10^{-6}$ Torr as well as a load-locked PLD system with a base pressure of $10^{-10}$ Torr. No systematic differences were observed between films grown in the two chambers. $Al_2O_3$ (R-plane and C-plane), $LaAlO_3$, and $SrTiO_3$ substrates were mounted on a heater plate using silver paint. The results for all of these oxide substrates were comparable.

Several different targets were prepared for this study. $MgB_2$ powders were either obtained by reacting stoichiometric amounts of elemental Mg turnings (99.99%) and amorphous B powder (< 45 microns) at 1000°C for several hours in a crimped Ta cylinder, or from a commercial source (Alfa Aesar). Ceramics with densities of about 65% were formed by sintering a 40 MPa cold-pressed disk at 890°C for 2 hours in a sealed tantalum/quartz tube or hot-pressing at 850°C with a pressure of 28 MPa in argon. Densities in excess of 90% were obtained by hot-pressing commercial $MgB_2$ powders at 1200°C for 1 hour. Ablation damage was strongly dependent on these processing parameters. However, no systematic dependence of film properties on these parameters was observed.

Figure 1 shows schematic drawings of the targets used and the structures grown in this work. As shown in Fig. 1a, a segmented target consisting of one half-disk of $MgB_2$ plus one half-disk of Mg has been employed, in a manner similar to earlier work on materials containing components of largely different volatility [9]. Single $MgB_2$ targets were also used, as was a single magnesium target to deposit Mg cap layers. The schematic cross-sections of the structures grown are shown in Fig. 1b.

Reasonable deposition rates were achieved with a background of $10^{-4}$ Torr of $Ar/4\%H_2$, laser energies ranging from 200 – 400 mJ per pulse at 248 nm, corresponding to a energy density of 1.7 – 3.3 $J/cm^2$ on the target, and a laser repetition rate of 15 Hz. Typical thicknesses of these films and Mg cap layers were 100 – 500 nm.

Somewhat surprisingly, a moderate increase in background pressure significantly decreased the deposition rate. Deposition experiments were performed with various gas pressures using both He and $Ar/4\%H_2$. Increasing the $Ar/4\%H_2$ pressure from vacuum to about 100 mTorr resulted in a change in plume color from bright green to faintly blue or purple (both for the Mg and the $MgB_2$ targets). At 200 mTorr $Ar/4\%H_2$ and a target-sample distance of 5 cm, no visible condensation onto the substrate at room temperature occurred (neither for the simple $MgB_2$ nor for the segmented target). Increasing the pressure to 600 mTorr, however, resulted in a powdery brown film even at T = 400°C. This may be due to a strong re-sputtering at 200 mTorr resulting from an energy transfer from the light B and/or Mg species to the much heavier Ar atoms. In fact, changing the background gas from 200 mTorr Ar/4%H$_2$ (where no growth occurs) to 200 mTorr of He resulted in the condensation of material on the substrate surface under otherwise identical conditions; however, the growth rates in this case were much poorer than at lower pressures.

Deposition experiments at elevated temperatures proved difficult due to the high vapor pressure of magnesium, even when using the segmented target or when the substrates were pre-coated with different starting layers, such as MgB$_2$, Mg, B, or Cu. Further experiments were performed in which the MgB$_2$ target was replaced by a hot-pressed pellet of boron (both as a single target and in a segmented target with Mg). However, the boron targets led to growth rates that were significantly lower than those from a MgB$_2$ pellet, required much larger laser energy densities, and were therefore not considered any further.

Films deposited from a MgB$_2$ or MgB$_2$/Mg segmented target at room temperature in 10$^{-4}$ Torr of Ar/4%H$_2$ were shiny with a metallic appearance. Within the accuracy of wavelength-dispersive spectroscopy (WDS), films grown from a MgB$_2$ target were stoichiometric but showed evidence for trace amounts of oxygen. Resistance measurements as a function of temperature revealed metallic conduction in these samples but no superconductivity down to 11K, indicating the need to post-anneal in order to form the desired MgB$_2$ phase.

To minimize magnesium loss during an *in-situ* annealing process, a Mg cap layer was deposited onto the as-deposited Mg-rich MgB$_2$ films (grown from the segmented target). The chamber was then filled with 0.7 atm. of Ar/4%H$_2$, and the samples were heated at a rate of about 100°C/min to various annealing temperatures ranging from 500°C to 650°C. Typical annealing time was 20 minutes, after which the sample was allowed to cool rapidly.

During the heating of the samples, it was observed visually that the color of the samples rapidly changed from bright silver to shiny dark at temperatures around 520°C to 550°C, at which point most of the excess magnesium appears to evaporate from the film. X-ray diffraction patterns for these samples (Fig. 2) confirm the presence of large amounts of unreacted magnesium for the film annealed at 500°C, with MgB$_2$ forming at and above 550°C (where elemental Mg is no longer detected in the x-ray patterns). The film annealed at 500°C (still containing a Mg cap layer) was further annealed at 630°C, and it was observed that the correct phase could no longer be obtained.

Figure 3 shows the resistance versus temperature curves for films subjected to single anneals at 500°C, 550°C, 600°C, as well as the sample annealed first at 500°C and subsequently at 630°C. Superconductivity is observed only for films annealed in the rather narrow temperature interval from 550°C to 600°C. The best films show a T$_c^{onset}$ ≈ 26.5K and a T$_c^{zero}$ ≈ 22.5K. Scanning electron microscopy (SEM) images show the presence of micron-sized droplets, as is often observed in PLD of metals. As seen in Fig. 4, however, atomic force microscopy (AFM) indicates that between these particulates, the films are dense with a reasonably smooth surface ( ≈ 1.5 nm rms for a 500 nm x 500 nm area). These images are in stark contrast to the results obtained in our laboratory by the *ex-situ* annealing of boron films [7], where well-defined large (100 nm) grains, and a rough, porous appearance are observed [7].

While superconducting films are easily obtained by *in-situ* annealing of stoichiometric or Mg-rich Mg+B mixed films, depositing a magnesium film by PLD on top of an e-beam deposited boron film, followed by annealing at 600°C did not yield superconducting $MgB_2$. This clearly indicates the advantage of forming $MgB_2$ from films of a homogeneous magnesium/boron mixture as opposed to using a magnesium/boron bilayer.

Conversely, e-beam evaporated boron films reacted *ex-situ* at 900°C in magnesium vapor typically show $T_c$'s above 38K [7], which is much higher than the $T_c$'s of films grown here. The same high-temperature annealing of a PLD-grown Mg+B layer yielded $T_c^{(zero)} = 22K$, comparable to the *in-situ* annealed films. This indicates that temperatures above 600°C do not result in further improvement of the material. The reason for the relatively low $T_c$ for PLD-grown samples is not clear. Strain in the samples can likely be ruled out as *ex-situ* post-annealed PLD films (a process in which $MgB_2$ is decomposed) show the same low $T_c$'s. Target contamination also appears to be an unlikely cause, as it was observed that the $T_c$ of a film grown from a single $MgB_2$ target and post-annealed at 900°C was significantly lower than that of the target itself. It may, however, be possible that contaminants in the target (e.g. MgO) segregate to grain boundaries in the ceramic where they have little influence on the $T_c$'s, while they are more uniformly distributed throughout the PLD-grown films. It is also possible that other Mg-B compounds ($MgB_4$, $MgB_6$, etc.) are formed in the energetic plume or on the target surface, and these compounds are too stable to be decomposed during the anneals even at 900°C [10]. Finally, the observation that both PLD systems with significantly different background pressures yield comparable films indicates that contamination resulting from residual oxygen in the growth chamber is likely not to be the leading cause of the decrease of the $T_c$'s.

Despite the fact that the present films show $T_c$'s that are considerably lower than those observed in bulk $MgB_2$ or *ex situ* reacted boron films, it is important to note that the samples prepared by the present method (i.e. the room temperature deposition of a Mg-rich homogeneous Mg/B mixture followed by an *in-situ* anneal) are never exposed to temperatures high enough to decompose $MgB_2$. Therefore, this work represents a significant step towards the growth of heterostructures, junctions, and possibly epitaxial films.

The authors would like to acknowledge L. Zhang for SEM analysis, W. Acree for assistance in WDS measurements, and X.X. Xi and D. Blank for valuable discussions. This research is sponsored by the U.S. Department of Energy under contract DE-AC05-00OR22735 with the Oak Ridge National Laboratory, managed by UT-Battelle, LLC,

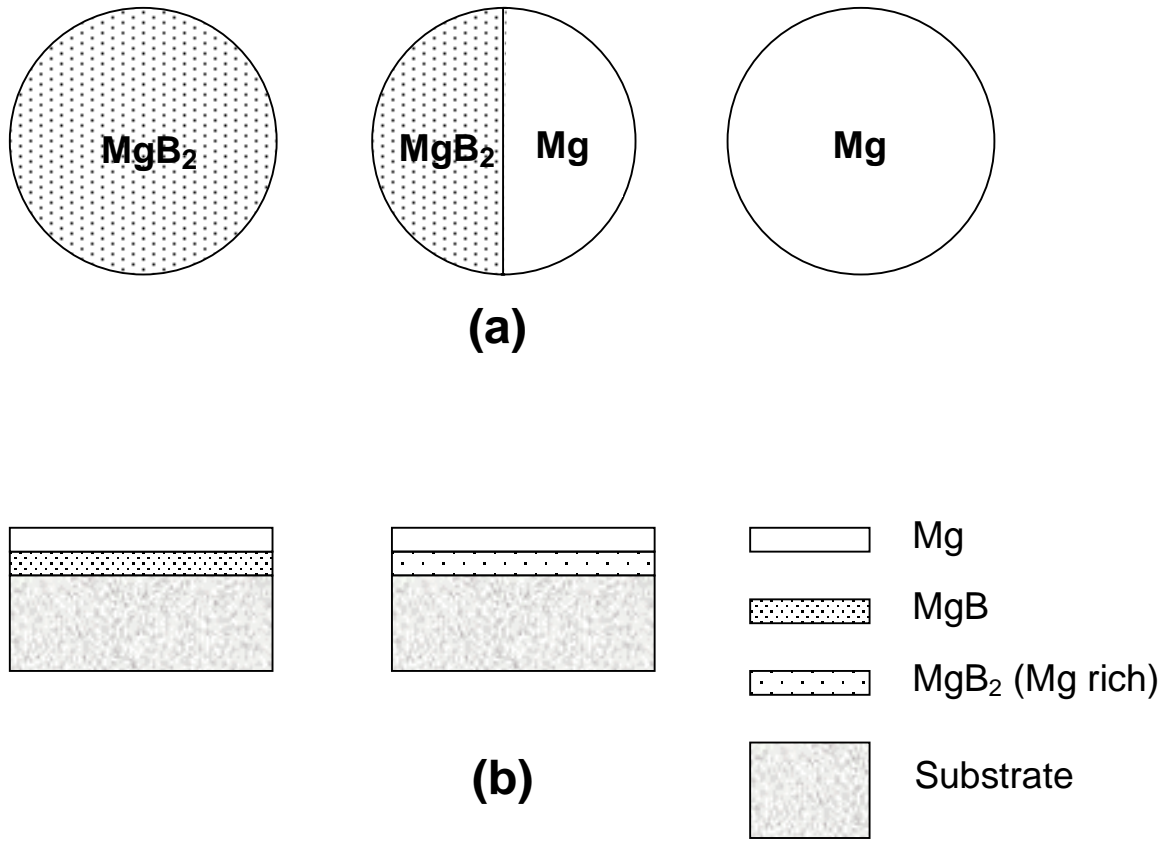

Figure 1. (a) Schematic drawings of the targets used for this work. Both $MgB_2$ and Mg single targets were used, as well as a segmented target consisting of a half-segment of $MgB_2$ and a half-segment of Mg (to compensate for the higher volatility of magnesium). (b) Schematic cross-section of the samples prepared. Nominally stoichiometric $MgB_2$ as well as Mg-rich Mg+B mixtures were used and a Mg cap layer was deposited to reduce the amount of magnesium loss during the *in-situ* anneal.

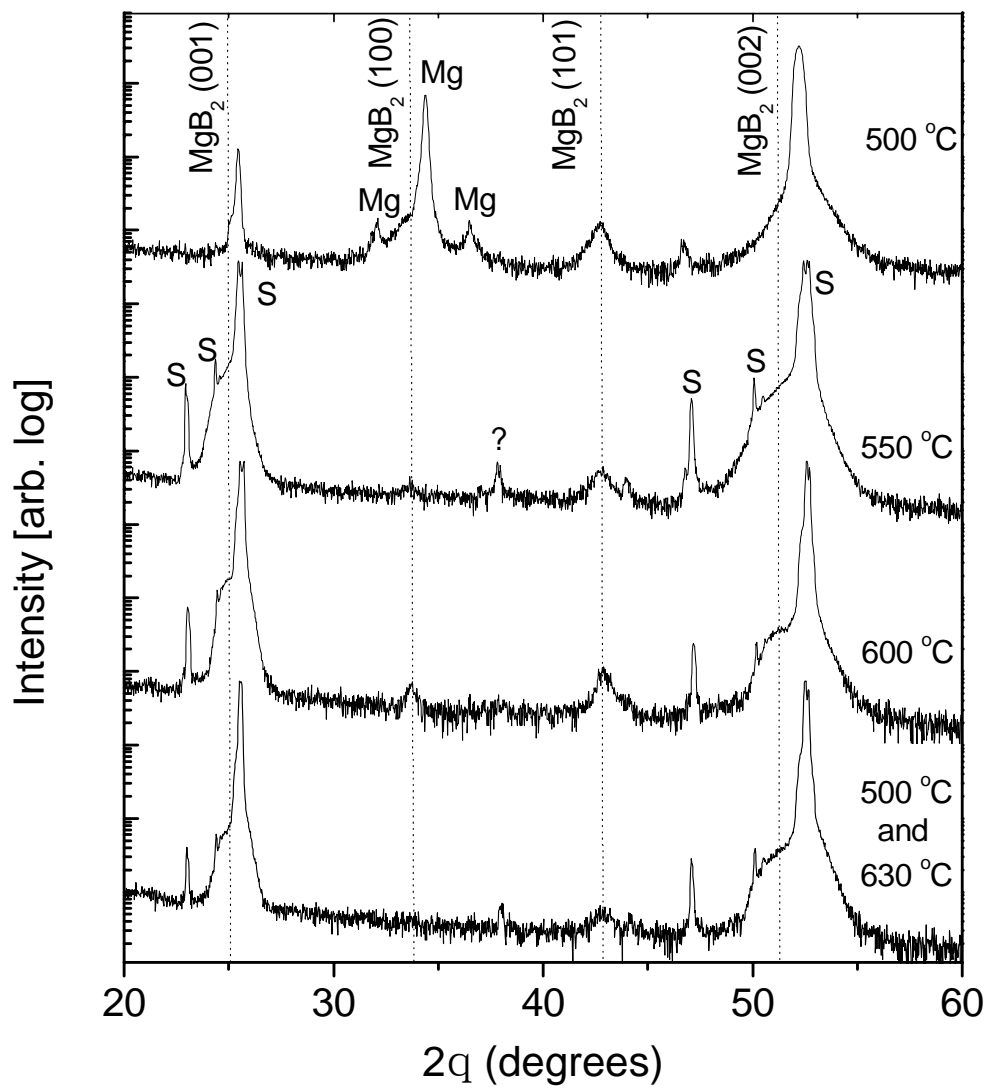

Figure 2. X-ray θ-2θ scans as a function of annealing temperature. The samples consisted of Mg-rich Mg+B mixture deposited onto R-plane $Al_2O_3$ and capped with a layer of Mg. Results for anneals at 500°C, 550°C, and 600°C are shown together with the result from the sample first annealed at 500°C and subsequently annealed at 630°C. Peaks corresponding to $MgB_2$ and Mg are labeled. Peaks labeled with "S" correspond to $Al_2O_3$ lines or spurious reflections and are also observed in patterns of a bare substrate.

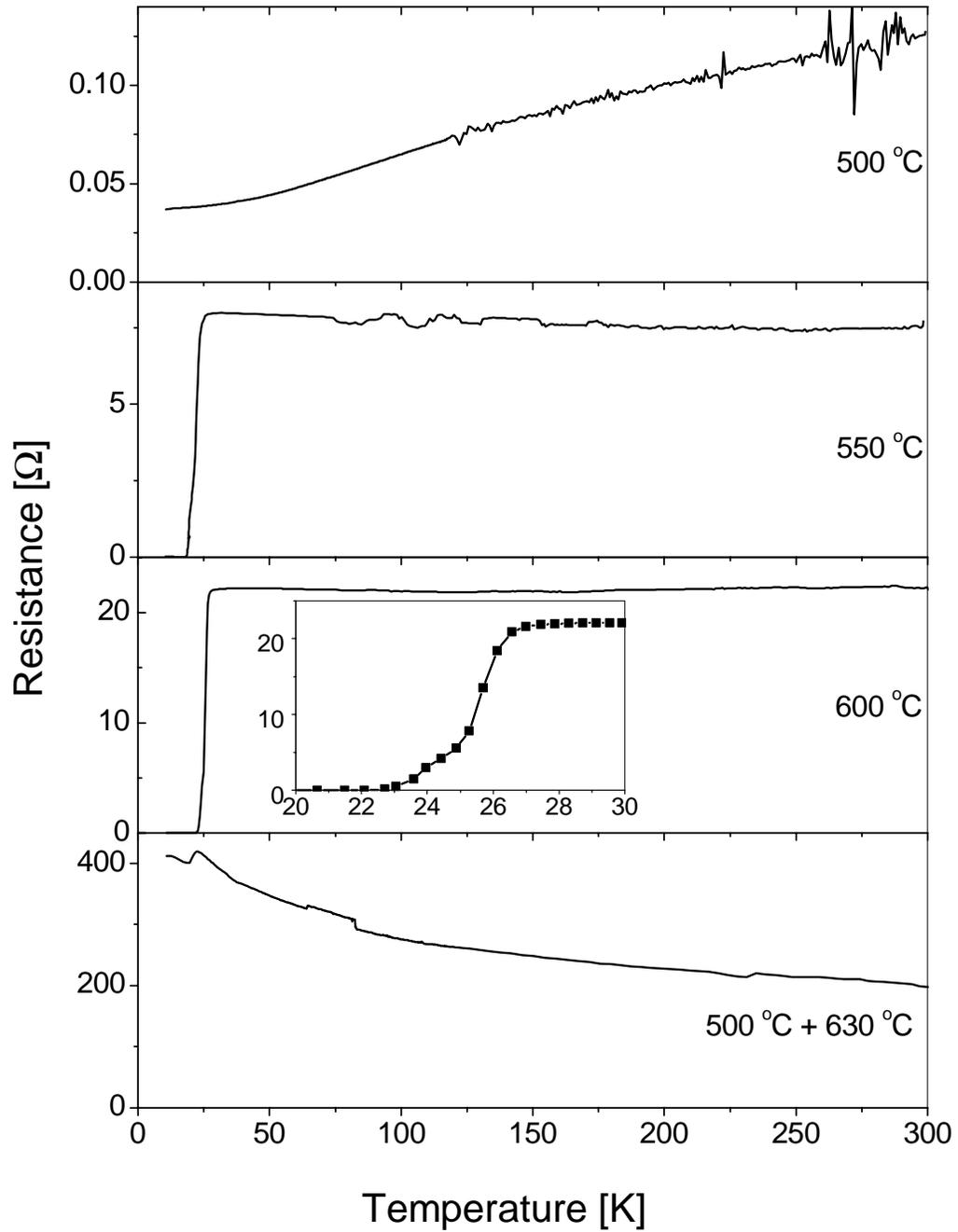

Figure 3. Resistance versus temperature curves for the same samples as in Fig. 2. The sample annealed at 600°C shows the highest transition temperature with $T_c^{onset} \approx 26.5K$ and $T_c^{zero} \approx 22.5K$.

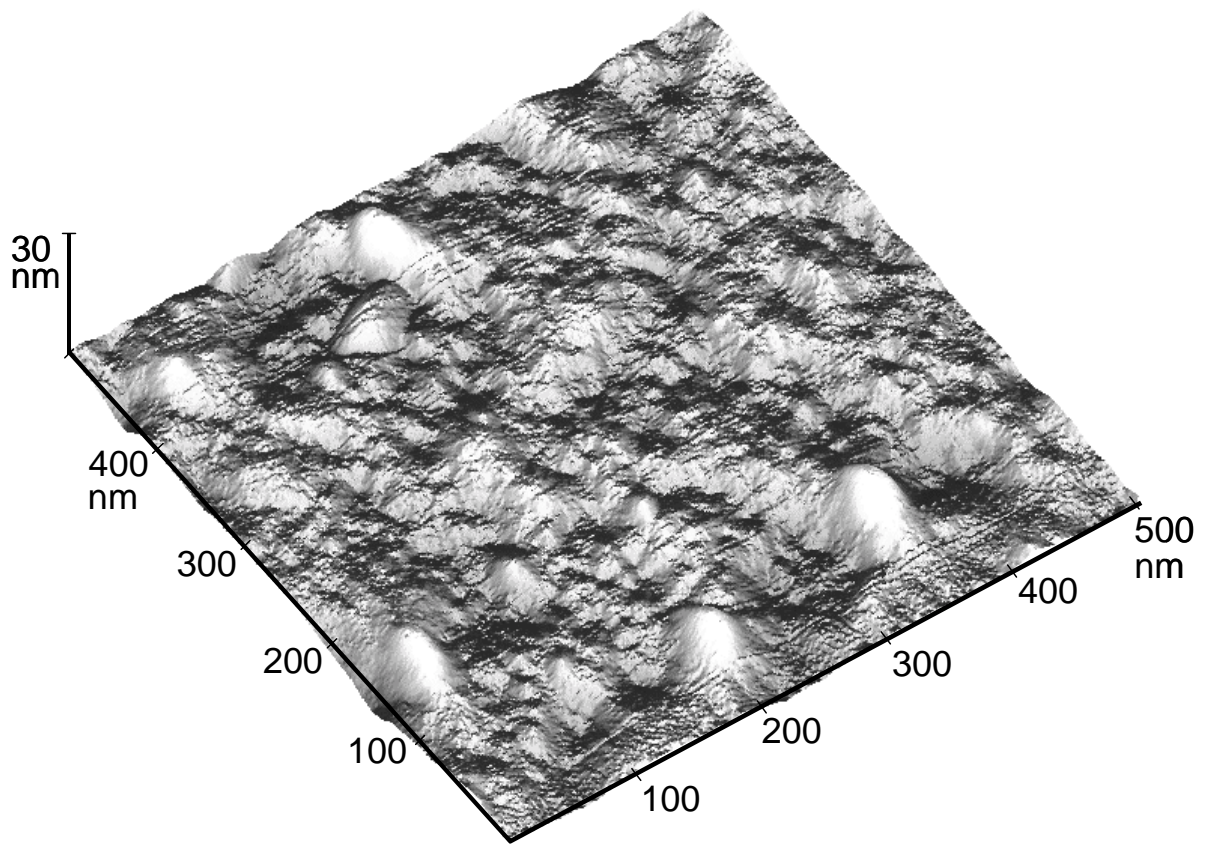

Figure 4. Atomic force micrograph of the sample annealed at 600°C. Except for the large droplets (not visible at this scale), which are assumed to have originated directly from the target, the films show a reasonably smooth and dense surface texture. Surface roughness for a 500 nm x 500 nm area chosen such as not to contain any of the large droplets is ≈ 1.5 nm (rms).